# Some phenomenological considerations on the nuclear collisions at high energies

I. V. Grossu [1], C. Besliu [1], Al. Jipa [1], D. Felea [2], C. C. Bordeianu [1]

[1] *University of Bucharest, Faculty of Physics, Bucharest-Magurele, Romania*
[2] *Institute of Space Sciences, Bucharest-Magurele, Romania*

We present some results obtained by applying the chaos theory on the numerical study of one three-dimensional, relativistic, many-body quark system. The asymptotic freedom property is introduced by employing a harmonic term in the bi-particle potential. In this context, we used also the outcome of a semiclassical study, applied to the quark constituents of nucleons. Depending on the initial temperature parameter, the system can evolve toward an oscillating or an expansion regime. It is important to notice also a transition region, characterized by a partial fragmentation (higher degree of order). This effect can be observed near the critical temperature and is related to the partial overcoming of the potential barrier (corresponding to the farthest particles from the system). The "degree of fragmentation" is defined on the Shannon entropy basis and using the graphs theory. For analyzing the expansion tendency of one relativistic many-body system, we employed also the "virial coefficient".

## 1. INTRODUCTION

One challenge of the last years in Physics, the Quark Gluon Plasma [1,2], brings the opportunity of reproducing, in laboratory, processes similar with the evolution of the early Universe at ages around one microsecond. The nucleus-nucleus collisions at relativistic energies are the only experimentally way to investigate this unusual state of the matter, which exists for a very short time interval (*5-10 Fm/c*), and which creation supposes high densities and temperatures ($\rho=2GeV/Fm^3$, $T=150-250MeV$).

In this paper we try to present an intuitive image of the complexity involved, by applying chaos theory to the evolution of a simplified, semi-classical, relativistic many-body system [3,4].

## 2. NUMERICAL METHOD

The simulation was implemented as an object oriented, C# application. The evolution of a many-body system is obtained by computing the following system of equations, for every moment of time:

$$\begin{cases} \vec{r}_i(t) = \vec{r}_i(t-dt) + \vec{v}_{mi} dt \\ \vec{p}_i(t) = \vec{p}_i(t-dt) - \nabla\left(\sum_j U_{ij}(r_{ij})\right) dt, \\ v = \dfrac{p}{\sqrt{m_0^2 + \dfrac{p^2}{c^2}}} \end{cases} \quad (1)$$

where, for each particle "$i$", $r_i$ is the position, $v_{mi}$ the average velocity for the infinitesimal interval $dt$, $p_i$ the momentum, $t$ is the time, $m_0$ the rest mass, $c$ the velocity of light in vacuum, and $U_{ij}$ the bi-particle potential.

The application can work also in a "retarded regime". By assuming that interactions propagates with the speed of light in vacuum, the evolution stages are stored in memory, and the interaction is considered only for the particle instances which have an appropriate temporal delay, with respect to their relative distance.

For analyzing the degree of stability [5,6], each simulation was implemented as a parallel processing of two identical systems, with slightly different initial conditions. Thus, based on the Lyapunov exponent method [7], we monitored the evolution in time of the following function:

$$L(t) \stackrel{def.}{=} \frac{1}{t} \ln \frac{d(t)}{d(0)}, \quad (2)$$

where $d(t)$ represents the distance in the phase space between the two mentioned systems, at the moment $t$ of time. For filtering the "geometrical variations" of $L(t)$, we considered only the points corresponding to a Poincare map [8].

Important information comes from the evolution in time of the system radius, defined as the average distance between all constituents. A framework based on Microsoft DirectX was also developed, for better representing the three-dimensional evolution of the n-body system.

### 3. STUDY ON QUARK MANY-BODY SYSTEMS

Inspired by the quark bag model [9], we chose a potential defined in three regions:

$$U_{ij} = \begin{cases} \dfrac{k}{2} r_{ij}^2 \;,\; r_{ij} \leq r_0 \\ \dfrac{q_i q_j}{4\pi\varepsilon\, r_{ij}} + \alpha r_{ij} \;,\; r \in (r_0, r_c) \;, \\ C, \quad r \geq r_c \end{cases} \quad (3)$$

where $r_{ij}$ is the distance between two quarks, $q$ the electric charge, $\varepsilon$ the permittivity of vacuum, $r_0$ an adjustable parameter, $r_c$ the confinement radius, $k$, $\alpha$ and $C$ are constants.

By developing the potential in Taylor series, the asymptotic freedom property is considered by using a harmonic term (Isgur-Karl model [10]). The second term introduces the coulombian interaction, together with the confinement, string potential [11]. For simplicity, the interaction is ignored for distances over the confinement radius.

For avoiding any geometrical effect, involved by the initial space distribution, each simulation was implemented as a homogenous mix of up and down quarks, placed in the vertices of a regular, centered dodecahedron.

As a quantum input, we considered an initial radial velocity distribution, calculated separately, for each quark flavor, in agreement with the average velocity resulted from the Fermi-Dirac statistics:

$$N(p)dp = \frac{4\pi V}{\hbar^3} p^2 \frac{g}{e^{\frac{E-\mu}{K_B T}} + 1} dp, \quad (4)$$

where $N(p)$ is the density of particles with momentum $p$, $V$ represents the volume occupied by the system, $g$ is the degeneracy level (for which we considered the color and the spin), $E$ the energy, $\mu$ the chemical potential, $T$ the temperature, $h$ the Plank constant, and $K_B$ represents the Boltzmann constant.

The chemical potential is determined with respect to the total number of constituents, given by:

$$N = \int N(p)dp. \quad (5)$$

Another quantum input comes from a semi-classical study [12], applied to the quark constituents of proton and neutron. For the particular case of a 3-body problem with harmonic bi-particle interaction, the classical closing trajectories condition (Lissajous) is:

$$\sqrt{\frac{3m_d}{2m_u + m_d}} \in \mathbf{Q}, \quad (6)$$

where $m_d$ and $m_u$ are the masses for the down and, respectively, up quarks, and $Q$ is the set of rational numbers. In the hypothesis of quark masses quantification, the previous equation leads to a relatively agreement with the known values of quark masses [13].

Our main goal was to find some phenomenological connection between the output obtained from the numerical analysis of one many-body system, and the formation of Quark Gluon Plasma. Thus, the initial conditions were chosen in agreement with the existing prediction for the hot, dense fireball evolution stage: dodecahedron edge *0.5Fm* and temperatures between *100* and *200MeV*. For the potential parameters (3), we worked with: *α=197MeV/Fm*, $r_c$=*1.2Fm*. The constants (*k* and *C*) result from the continuity condition. The adjustable parameter $r_0$ was set to *0.7Fm* and, for the computations accuracy, we worked at a temporal resolution *dt=$10^{-6}$Fm/c*. The simulation time was limited to *100Fm/c*.

In the mentioned conditions, one can notice a decreasing (Fig.1) of the Lyapunov Exponent Eq. (2), for temperatures over a critical value (close to *Tc=195MeV*). We tried to relate, intuitively, this value with the critical temperature corresponding to the phase transition to QGP, and the time interval (≈*4Fm/c*) until the Lyapunov Exponent reaches its saturation value, with the lifetime of QGP. The evolution in time of the average distance between constituents (Fig.2) shows the existence of an oscillating regime for temperatures bellow the critical value, followed by a slow increase (this last effect is related to the "evaporation" of some constituents). The evolution for temperatures over the critical value is, practically, linear.

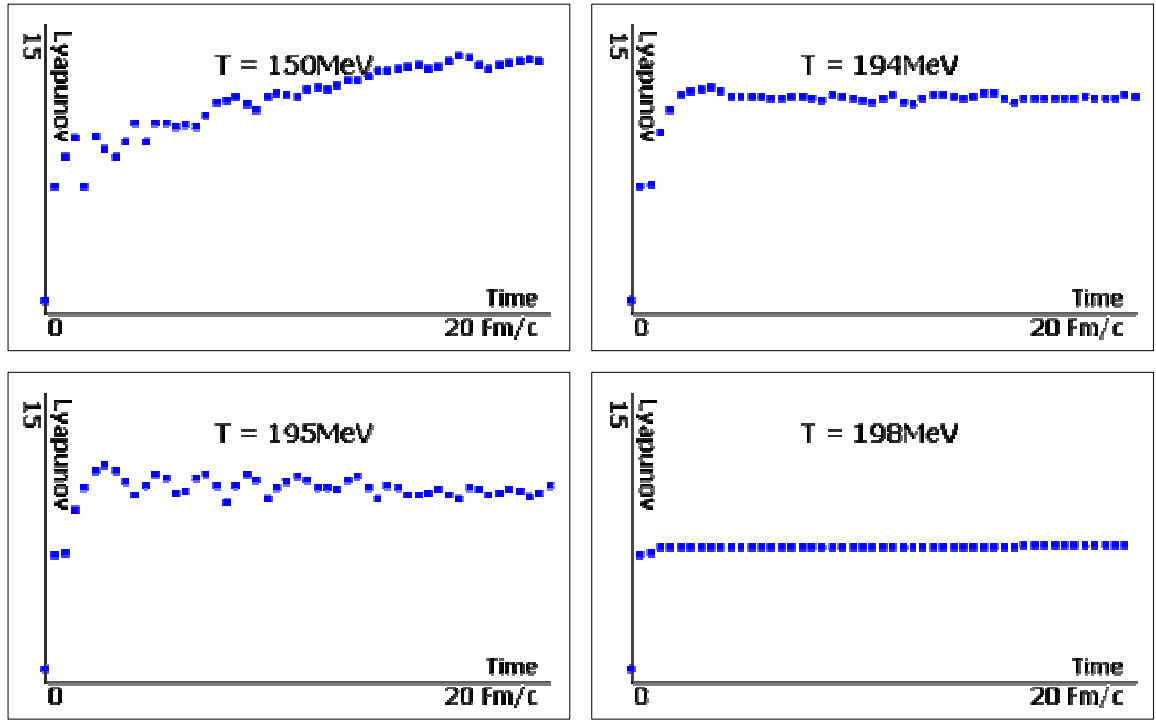

Fig.1 TheLyapunov Exponent for several temperatures, below and above the critical value (≈195MeV).

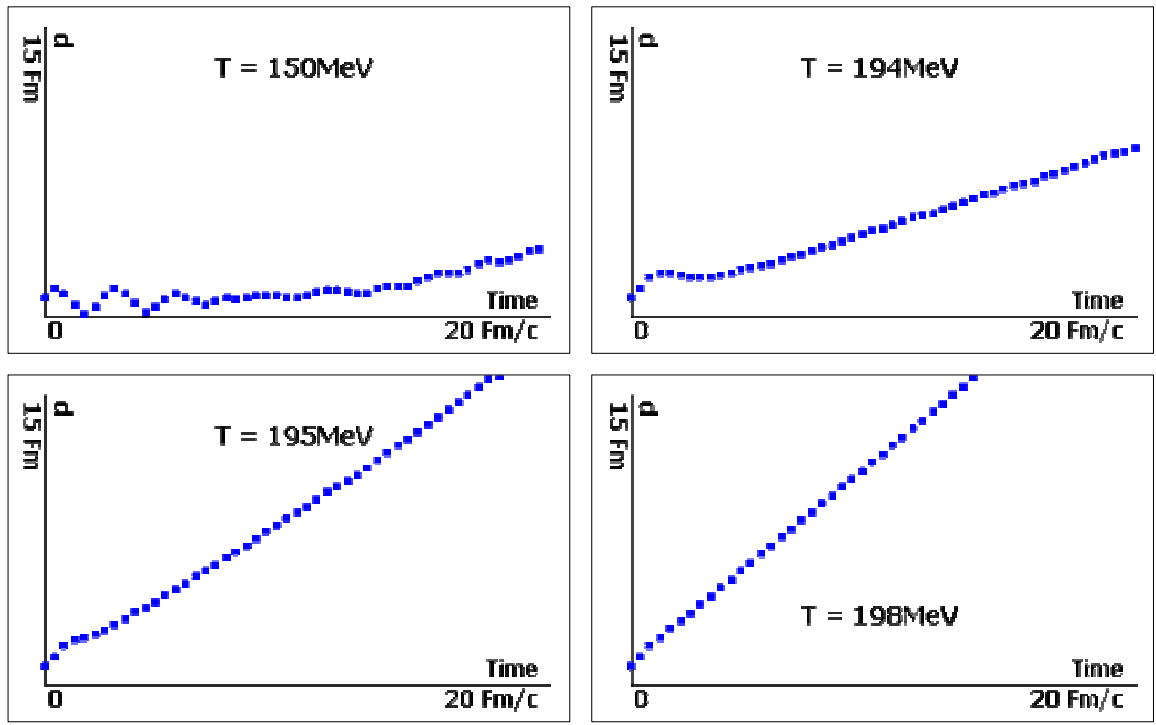

Fig.2 The system radius for several temperatures, below and above the critical value (≈195MeV).

## 4. THE ANALYSIS OF CLUSTERS

We first tried to define the concept of cluster. One can consider the graph $G(P,U)$ corresponding to a many body system, where $P$ represents the set of all constituents, and U is the set of all particle pairs for which the distance is lower than the interaction radius. Thus, the set of clusters could be linked to the set of all maximal distinct connex sub-graphs of $G(P,U)$.

The degree of fragmentation could be defined starting from the Shannon entropy:

$$F = \sum_{i=1}^{N_C} f_i \ln f_i; \quad f_i = \frac{n_i}{n} \quad (7)$$

where $N_C$ is the number of clusters, $n_i$ the number of constituents for each cluster, and $n$ is the total number of constituents of the system.

A global information on the fragmentation could be obtained by considering the superposition of the Poincare maps corresponding to the functions $p_x=p_x(x,y)$, $p_y=p_y(X,Y), p_z=p_z(X,Y)$, for all the particle of the n-body system. As expected, we noticed that the fractal dimension of the global Poincare maps decreases with the degree of fragmentation $F$ (Eq.7).

For analyzing the expansion tendency of one relativistic many-body system we considered the following quantity (the "virial coefficient"):

$$V = \lim_{t \to \infty} \frac{\sum_{k=1}^{n} <p_k v_k>}{\sum_{k=1}^{n} <F_k r_k>} \quad (8)$$

where $p_k$ is the momentum, $v_k$ the velocity, $r_k$ the position, and $F_k$ the total force acting on each particle "$k$". The bracket denotes the average over the time $t$.

One advantage of the virial theorem [15] comes from it's highly degree of generality. It does not depend on the notion of temperature and holds even for systems that are not in thermal equilibrium. Thus, the absolute value of $V$ (Eq.8) is unity for a stable, bound system, and increases with the expansion tendency. A value less than unity could be connected with the system's collapse.

5. CONCLUSIONS

In the attempt of finding some connections between the evolution of a simplified, semi-classical, relativistic many-body system and the formation of QGP, we considered some intuitive hypotheses:
1. based on the quark bag model [9], we considered a bi-particle potential Eq. (3), defined in three regions;
2. the asymptotic freedom property was implemented through a harmonic potential term;
3. for avoiding geometrical effects, we worked with a homogenous mix of up and down quarks, initially placed in the vertices of a regular, centered dodecahedron;
4. one quantum input comes from using initial radial velocities in agreement with the average value resulted from the Fermi-Dirac statistics (4);
5. another quantum input comes from a semi-classical study [12], applied to the quark constituents of proton and neutron.

We analyzed the evolution of a particular class of isolated systems, far from equilibrium, and for which the interactions between constituents play a significant role. Depending on the initial temperature parameter, the system can evolve toward an oscillating or an expansion regime. It is important to notice also a transition region, characterized by a partial fragmentation (higher degree of order). This effect can be observed near the critical temperature and is related to the partial overcoming of the potential barrier (corresponding to the farthest particles from the system).

Some encouraging results are related to the critical temperature (≈195MeV), which is in a good agreement with the QCD calculations *Tc=192MeV* [14], and to a relatively good value *(≈4Fm/c)* for the QGP life time (*5-10Fm/c* in accordance with estimates based on collision energies reached at RHIC).

As discussed in the section 2, the analysis of a retarded interaction imposes some computational difficulties (long processing time and memory intensive). Our first investigations indicate, in this case, an increase of instability (higher values for the Lyapunov exponent Eq. (2)), and a lower critical temperature.

One can notice also the high dependence on initial conditions (positive values for the Lyapunov Exponent (Fig.1)). Thus, we have a better vision on the important role played by the geometry in nuclear relativistic collisions.

On the other hand, we observed the importance of the dimensionality. As opposed to the two-dimensional case, where we dispose of infinity regular polygons, for a three-dimensional system, we are limited to only five regular polyhedrons, which fact could be related with a clusterization tendency.